\documentclass[twocolumn,showpacs,preprintnumbers,amsmath,amssymb, prl]{revtex4}

\usepackage{graphicx}

\begin{document}

\title{\bf Continuous Time Quantum Monte Carlo Method for
Fermions:\\ Beyond Auxiliary Field Framework}

\author{A. N. Rubtsov}
\email{alex@shg.ru} \affiliation{Department of Physics, Moscow
State University, 119992 Moscow, Russia }
\author{A. I. Lichtenstein}
\email{a.lichtenstein@sci.kun.nl} \affiliation{Institute of
Theoretical Physics, University of Hamburg, Jungiusstrasse 9,
20355 Hamburg, Germany} 

\begin{abstract}
Numerically exact continuous-time Quantum Monte Carlo algorithm
for finite fermionic systems with non-local interactions is
proposed. The scheme is particularly applicable for general
multi-band time-dependent correlations since it does not invoke
Hubbard-Stratonovich transformation. The present determinantal
grand-canonical method is based on a stochastic series expansion
for the partition function in the interaction representation. The
results for the Green function and for the time-dependent
susceptibility of multi-orbital super-symmetric impurity model
with a spin-flip interaction are presented.
\end{abstract}



\maketitle

Quantum Monte Carlo (QMC) tools for fermionic systems appeared
more than 20 years ago \cite{Scalapino, Blankenbecler, Hirsch,
HirschPH} and are nowdays vital for a wide range of fields, like
the physics of correlated materials, quantum chemistry and
nanoelectronics. Although the first programs were developed for
model Hamiltonians with local interaction, many-particle action of
a very general form stays behind the real systems. For example all
matrix elements of the interaction do not vanish in the problems
of quantum chemistry \cite{qchem} and solid state physics
\cite{qmcpw}. Dynamical mean-field theory (DMFT) \cite{DMFT} for
correlated materials brings a non-trivial bath Green function on
the scene, and its extension \cite{TDU} deals with an interaction
which is non-local in time. An off-diagonal exchange term can be
responsible for the correlated superconductivity in doped
fullerens \cite{Fullerens}. It is worth to note in general that
exchange is often of an indirect origin (like super-exchange) and
the exchange terms are therefore retarded. New developments
\cite{newDMFT} clearly urge an invention of essentially different
type of QMC scheme suitable for non-local, time-dependent
interaction.

The determinantal grand-canonical auxiliary-field scheme
\cite{Scalapino, Blankenbecler, Hirsch, HirschPH} is commonly used
for the interacting fermions, because other known QMC schemes
(like stochastic series expansion in powers of Hamiltonian
\cite{SSE} or worm algorithms \cite{Worm}) suffer an unacceptably
bad sign problem for this case. Two points are essential for the
approach: first, the imaginary time is artificially discretized,
and then the Hubbard-Stratonovich transformation \cite{Hubbard} is performed to
decouple the fermionic degrees of freedom. After the decoupling,
fermions can be integrated out, and Monte Carlo sampling should be
performed in the space of auxiliary Hubbard-Stratonovich fields.
Hirsh \cite{Hirsch} proposed to use discrete Hubbard-Stratonovich
transformation to improve the original scheme; this is now a
standard method for simulations of lattice and impurity quantum
problems. For relatively small clusters, and in particular for
DMFT, the sign problem is not crucial in this method
\cite{clusterDMFT}. The number of auxiliary field is linear
(quadratic) in the number of atoms for the case of local
(nonlocal) interaction.

The time discretization leads in a systematic error of the result.
For for bosonic quantum systems, continuous time loop algorithm
\cite{Beard}, worm diagrammatic world line Monte Carlo scheme
\cite{Worm} and continuous time path-integral
QMC\cite{Kornilovitch} overcame this issue. Recently a
continuous-time modification of the fermionic QMC algorithm was
proposed \cite{CT}. It is based on a series expansion for the
partition function in the powers of interaction. The scheme is
free of time-discretization errors, but the Hubbard-Stratonovich
transformation is still invoked. Therefore the number of auxiliary
fields scales similarly to the discrete scheme. This scheme is
developed for local interaction only.

Besides the time-discretization problem, the non-locality of
interaction hampers the calculation in the existing schemes,
because it is hard to simulate systems with a large number of
auxiliary spins. Further, the discrete Hubbard-Stratonovich
transformation is not suitable for non-local in time interactions.
One needs to use continuous dispersive bosonic fields
\cite{TDU} for this case, that makes the simulation even
harder.

In this Letter we present a novel numerically exact
continuous-time fermionic QMC algorithm. This is the first QMC
scheme that do not invoke any type of Hubbard-Stratonovich
transformations and therefore operates natively with non-local in
space and time interactions. The scheme is free of systematic
errors due to direct operations with continuous time expansion of
the partition function. Numerical results for a super-symmetric
two band impurity model with spin-flip, time dependent non-local
interactions show an advantage and a broad perspective of proposed
QMC scheme for the complex solid-state and quantum chemistry
problems.

We consider a fermions system with pair interaction in the most
general form and present the partition function $Z={\mathrm{Tr}
\hspace{0.1em}}T\exp (-S)$ in the terms of the effective action
$S$:
\begin{eqnarray}
\label{Z0}
    &S=S_{0}+W\equiv \int \int   t_r^{r'}  c^\dag_{r'} c^{r} dr
    dr'+ \\ \nonumber
    &\int \int \int \int
     w_{r_1 r_2} ^{r_1' r_2'} (c^\dag_{r_1'} c^{r_1}-\alpha^{r_1}_{r_1'})
    (c^\dag_{r_2'} c^{r_2}-\alpha^{r_2}_{r_2'}) dr_1 dr_1' dr_2 dr_2'.
\end{eqnarray}
Here $T$ is a time-ordering operator, $r=\{i,s,\tau\}$ is a
combination of the discrete index $i$ numbering the
single-particle states in a lattice, spin index $s=\uparrow$ or
$\downarrow$ and the continuous imaginary-time variable $\tau$.
Integration over $d r$ implies the integral over $d\tau$, and the
sum over all lattice states and spin projections: $\int dr \equiv
\sum_i \sum_s \int_0^\beta d\tau$. We borrow the linear-algebra
style for sub- and superscripts to make the notation clearer. The
creation ($c_{r}^{\dag })$ and annihilation $(c^{r})$ operators
for a fermion in the state $r$ are labelled as covariant and
contravariant vectors, respectively. The labelling for
coefficients $t, w$ is chosen to present all integrands like
scalar products of tensors. An additional quantity $\alpha
_{r^{\prime }}^{r}$ is introduced for the most effective splitting
of $S$ to the Gaussian part ($S_{0}$) and interaction ($W$). The
parameters $\alpha _{r^{\prime }}^{r}$ are to be chosen later to
to optimize the algorithm and to minimize the sign problem.

We consider $S_{0}$ as an unperturbed action and switch to the interaction
representation. The perturbation-series expansion for $Z$ has the following
form:
\begin{eqnarray}\label{ser}
    &Z=\sum_{k=0}^\infty \int dr_1 \int d r_1' ...  \int
    dr_{2k}' \Omega_k (r_1, r_1', ...,  r_{2k}')\\
     \nonumber
     &\Omega_k=Z_0 \frac{(-1)^k}{k!}w_{r_1 r_2}^{r_1' r_2'} \cdot ... \cdot w_{r_{2k-1} r_{2k}}^{r_{2k-1}' r_{2k}'}
     D^{r_1 r_2 ... r_{2k}}_{r_1' r_2' ... r_{2k}'}
\end{eqnarray}
where $Z_0$ is a partition function for the unperturbed system and
\begin{equation}\label{D}
    D^{r_1 ... r_{2k}}_{r_1' ... r_{2k}'}
    = <T (c^\dag_{r_1'} c^{r_1}-\alpha^{r_1}_{r_1'}) \cdot ... \cdot (c^\dag_{r_{2k}'} c^{r_{2k}}-\alpha^{r_{2k}}_{r_{2k}'})>.
\end{equation}
Hereafter the triangle brackets denote the average over the
unperturbed system, $<A>=Z_{0}^{-1}\mathrm{Tr} \hspace{0.1em}\{T A
\exp (-S_{0})\}$. Since the action $S_{0}$ is Gaussian, one can
apply the Wick theorem and find the expression for $D$ in terms of
a determinant of $2k\times 2k$ matrix:
\begin{equation}\label{det}
    D^{r_1 r_2 ... r_{2k}}_{r_1' r_2' ... r_{2k}'}=
    \det|| g^{r_i}_{r_j'}-\delta _{ij} \alpha^{r_i}_{r_j'} ||
\end{equation}
Here $g_{r'}^r=<T c^\dag_{r'} c^{r}>$ is the the single-particle
two-point Green function in the QMC notation and $\delta _{ij}$ is
a delta-symbol.

In the following we use an important-sampling Markov process in
the configuration space, where the points are determined by the
perturbation order $k$ and the set $\{r_1, r_1', ...,r_{2k}'\}$.
Suppose for a moment that $\Omega$ is always positive, and
consider a random walk with a probability of $Z^{-1} \Omega_k(r_1,
r_1', ..., r_{2k}, r_{2k}')$ to visit each point. Denote the
average over this random walk by the overline. Then for example
the Green function $G_{r'}^r=Z^{-1} <T c^\dag_{r'} c^{r} e^{-W}>$
can be expressed as $\overline{g_{r'}^r (r_1, r_1', ...,
r_{2k}')}$ where $g_{r'}^r$ determines the Green function for a
current realization. It is important to note that a Fourier
transform of $g_{r'}^r$ with respect to time arguments can be
found analytically. Therefore the Green function can be calculated
directly at Matsubara frequencies. Such an approach has an
advantage over the calculation in $\tau$-domain, because it
automatically takes into account the invariance of the initial
action in the translations along $\tau$-axis. Higher-order
correlators can be calculated in the same way. More detailed
description of the algorithm as well as methodological discussion
can be found in Ref. \cite{AR}.

In certain cases proper choice of $\alpha$ can indeed completely
suppress the sign problem. For example, for Hubbard model it is
reasonable to choose $\alpha _{i^{\prime }s^{\prime } t^{\prime
}}^{i s t}=\alpha _s \delta _{\tau \tau ^{\prime }}\delta _{ij}
\delta_{s s^{\prime }}$. If the Gaussian part of action does not rotate
spins, than $g^s_{s^{\prime}}\propto \delta_{s s^{\prime }}$, and
the determinant in (\ref{det}) is factorized: $D=D_\uparrow
D_\downarrow$. For the case of Hubbard model with attraction one
should choose $\alpha _{\uparrow }=\alpha _{\downarrow }=\alpha$,
where $\alpha $ is a real. For this choice
$g_{\downarrow}^{\downarrow}=g_{\uparrow }^{\uparrow }$, and
consequently $D_{\uparrow }=D_{\downarrow }$. All terms of are
positive in this case, because $w<0$.

The choice of $\alpha _{\uparrow }=\alpha _{\downarrow }$ is
useless for a system with repulsion, because the alternating signs
of $\Omega _{k}$ with odd and even $k$ appear \cite{Alt}.
Similarly to the discrete Hubbard-Stratonovich transformations
\cite{HirschPH}, the particle-hole symmetry can be exploited for a
half-filled system. One can show that a choice $\alpha _{\uparrow
}=1-\alpha _{\downarrow }=\alpha$ delivers a condition
$D_{\uparrow }=-D_{\downarrow }$ for this case, thus eliminating
the sign problem \cite{AR}. Further, for a particular case of an
impurity problem in the atomic limit $\alpha _{\uparrow }=1-\alpha
_{\downarrow }=\alpha$ with $\alpha>1$ or $\alpha<0$ eliminates
the sign problem for the repulsive interaction at any filling
factor \cite{AR}.

Summarizing up these observations, we can write a draft recipe of
how to choose $\alpha$. For a physically reasonable split of the
action \ref{Z0} the value of $\alpha$ should not be too large.
Therefore for the diagonal repulsive terms of the interaction
matrix we propose to use $\alpha _{\uparrow }=1-\alpha
_{\downarrow }=\alpha$ with $\alpha$ slightly above 1. For the
attractive interaction, and for the off-diagonal matrix elements
of $w$, the choice should be $\alpha _{\uparrow
}=\alpha_{\downarrow} \approx 0.5$. Of course, in a general case
$\Omega$ is not positive-defined and one needs to work with its
absolute value in QMC sampling. In this case an exponential
fall-off occurs for the large systems or small temperature. It is
worth while to mention that an above choice of parameters $\alpha$
suppress the sign problem for local DMFT-like action with diagonal
in orbital indices bath Green function.

Now we discuss how to organize a random walk in practice. We need
to perform a random walk in the space of $k; r_1, r_1', ...,
r_{2k}'$. Two kinds of trial steps are necessary: one should try
either to increase or to decrease $k$ by 1, and, respectively, to
add or to remove the four corresponding operators. A proposition
for $r_{2k+1}, r'_{2k+1}, r_{2k+2}, r'_{2k+2}$ should be generated
for the "incremental" step. The normalized modulus
\begin{eqnarray}
&||w||^{-1} |w_{r_{2k+1} r_{2k+2}}^{r_{2k+1}' r_{2k+2}'}| \\
\nonumber &||w||={\int \int \int \int  |w_{r R}^{r' R'}|} dr dR
dr' dR'
\end{eqnarray}
can be used as a probability density for this proposition. Then
the standard Metropolis acceptance criterion can be constructed
using the ratio
\begin{equation}\label{prob}
    \frac{||w||}{k+1} \cdot
    \left| \frac{D^{r_1  ... r_{2k+2}}_{r_1' ... r_{2k+2}'} }
    {D^{r_1  ... r_{2k}}_{r_1' ... r_{2k}'} } \right|.
\end{equation}
The "decremental" step can be organized in a same way.

The most time consuming operation of the algorithm is a
calculation of the ratio of determinants, defined by the Eq.
(\ref{det}). Fast update trick can be used, resulting in $\propto
k^2$ operations \cite{Scalapino, HirschPH}. Here we estimate $k$. An average value of
(\ref{prob}) determines an acceptance rate for QMC sampling. It is
reasonable to expect that by the order of magnitude this rate is
not much less then unity. The ratio of determinants times $||w||$
can be interpreted as an expectation value for $|W|$. Therefore
\begin{equation} \label{Estk}
    k \approx \overline {|W|}.
\end{equation}
For the Hubbard lattice of $N$ atoms with an interaction constant
$U$, for instance, $|W| \propto \beta |U| N$. In principle, one
can manipulate with $\alpha$ to minimize $\overline {|W|}$. These
manipulations should however preserve the average sign as large as
possible.
\begin{figure}
\includegraphics[width=7cm]{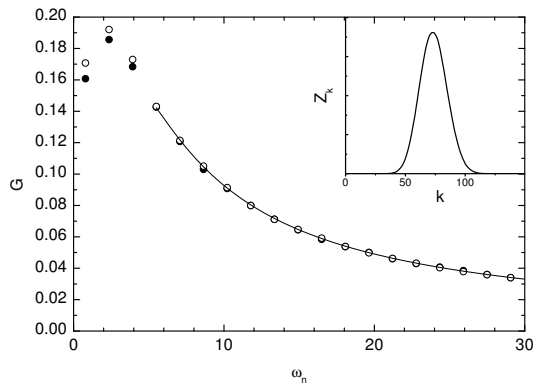}%
\caption{ \label{fig:epsart}Local Green function of the two-band
rotationally invariant model at the Matsubara frequencies. Filled
and open circles correspond to the static and to the nonlocal in
time spin-flip, respectively. High-frequency asymptotics is drawn
with line. Inset shows the distribution function for the
perturbation
order $k$.} %
\end{figure}
\begin{figure}
\includegraphics[width=7cm]{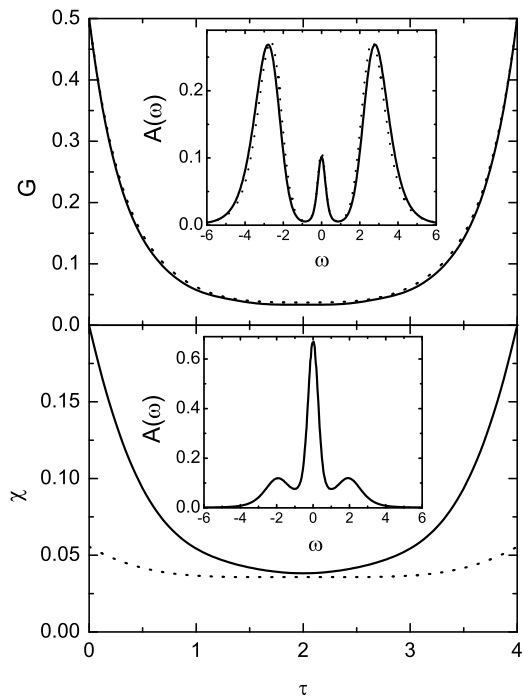}
\caption{\label{fig:epsart} Imaginary-time Green Function (upper
panel) and the four-point correlator $\chi$ (lower panel) for two
band model. Upper inset shows DOS computed from the Green
function. Solid and dot lines correspond to the static and to the
nonlocal in time spin-flip, respectively. Lower inset shows DOS
for 5 band model with the same value of U and J=0.2}
\end{figure}

We apply the proposed continuous time QMC for the important
problem of super-symmetric two band impurity model at half-filling
\cite{Rozenberg, Dworin}. To our knowledge, this is the first
successful attempt to take the off-diagonal exchange terms of this
model into account. These terms are important for the realistic
study of multi-band Kondo problem, because they are responsible
for the local moment formation \cite{Dworin}. The interaction in
this model has the following form
\begin{equation}
\frac{U}{2} (\hat{N}(\tau)-2) (\hat{N}(\tau)-2) - \frac{J}{2}
({\bf S}(\tau) \cdot {\bf S}(\tau)+ {\bf L}(\tau) \cdot {\bf
L}(\tau)),
\end{equation}
where $\hat{N}$ is the operator of total number, $S$ and $L$ are
total spin and orbital-momentum operators, respectively. The
interaction is spin- and orbital- rotationally invariant. The
Gaussian part of the action represents the diagonal semicircular density of
states \cite{DMFT} with unitary half-band width: $t(\omega
)=2/(\omega_n +\sqrt{\omega_n^{2}-1})$, where $\omega_n$ are
Matsubara frequencies related to imaginary time variable. We used
parameters $U=4, J=1$ at $\beta=4$. A modification of this model
was also studied, where spin-flip operators were replaced with the
fully non-local in time terms. For example, operator
$c^{\dagger}_{0 \uparrow \tau} c^{0 \downarrow \tau}
c^{\dagger}_{1 \downarrow \tau} c^{1 \uparrow \tau}$  was replaced
with $\beta^{-1}\int d\tau^{\prime} c^{\dagger}_{0 \uparrow \tau}
c^{0 \downarrow \tau} c^{\dagger}_{1 \downarrow \tau^{\prime}}
c^{1 \uparrow \tau^{\prime}}$. Figures present the result for the
local Green function $G_{i s}^{i s}$ and the four-point correlator
$\chi(\tau-\tau^{\prime})= <c^{\dagger}_{0 \uparrow \tau} c^{0
\downarrow \tau} c^{\dagger}_{1 \downarrow \tau^{\prime}} c^{1
\uparrow \tau^{\prime}}>$. The later quantity characterizes the
spin-spin correlations and would vanish if the exchange is absent.

Figure 1 shows the Green function at Matsubara frequencies. The
typical number of QMC trials was $2 \cdot 10^7$. Results for the
local and non-local in time spin-flip are shown with filled and
open circles, respectively. The distribution function for the
perturbation order $k$ is drawn in the inset. For the system
studied it appears to be a Gaussian-like peak located at $k\approx
75$, with an acordance to Eq. (\ref{Estk}). The estimated error
bar in $G(\omega_n)$ is about $3 \cdot 10^{-3}$ for the lowest
frequency and becomes smaller as the frequency increases. The
high-frequency tail obeys an asymptotic behavior $-{\rm Im} (i w+
\epsilon)^{-1}$ with $\epsilon \approx 2.9$.

Green function in the time domain was obtained by a numerical
Fourier-transform from the data for $G(\omega_n)$. For high
harmonics the above-mentioned asymptotic was used. Results are
presented in the upper panel of Fig.2. The lower panel presents
the result for $\chi(\tau)$. Thess data are obtained similarly,
the difference is that $\chi(\omega)$ is defined at Bose Matsubara
frequencies and obeys a $1/\omega^2$ decay. It is interesting to
note that Green function is almost insensitive to the details of
spin-flip retardation. Both Green functions are very similar and
correspond to qualitatively the same density of states (DOS). The
maximum-entropy guess for DOS is presented in the inset to Fig.2.
On the other hand, switch to the non-local in time exchange
modifies $\chi(\tau)$ dramatically. The local in time exchange
results in a pronounces peak of $\chi(\tau)$ at $\tau \approx 0$,
whereas the non-local spin-flip results in almost time-independent
spin-spin correlations. For realistic description of Kondo
impurities like cobalt atom on metallic surface it is of crucial
importance to use the spin and orbital rotationally invariant
Coulomb vertex in the non-perturbative investigation of
electronic structure. The proposed continuous time QMC scheme is
easily generalized for a general multiband case. As example we
shows the DOS for five d-orbital model at half-filling for the
same value of U and J=0.2 in the lower insert of Fig.2.

For a final discussion it is suitable to analyze a convergence of
the series (\ref{ser}). Fermi statistics and a finite size of the
system insure us that the configurational space of the problem is
of a finite order. Because the perturbation operator $W$ has a
finite norm, its powers $W^k$ therefore grow slower than $k!$.
Consequently, from the mathematical point of view there is no
doubt that the series (\ref{ser}) always converges. Physically it
is important to note that this convergence is related both with a
choice of the type of serial expansion and with the peculiarities
of the system under study. First of all, series (\ref{ser})
contains {\it all} diagrams, including non-bounded. In the
analytical diagram-series expansion non-bounded diagrams drop out
from the calculation \cite{IZ}, and the convergence radius for the
diagram-series expansion differs from that of (\ref{ser}).
Further, Fermi statistics is indeed important. An analog of
(\ref{ser}) for Bose field can diverge even for a single-atom
problem \cite{IZ}, because in this case one deals with an
infinite-order Gilbert space. It is important to keep this in mind
for possible extensions of the algorithm to the electron-phonon
systems and to the field models, as these systems are also
characterized by an infinite-order phase space. A general
time-dependent form of the action (Eq. (\ref{Z0})) allowed us to
use renormalization theory for the Hubbard-like model: in this
case local DMFT  would be a starting point for lattice
calculations in order to reduce the effective interaction and
minimize the sign problem.

 In conclusion, we have developed a fermionic continuous time quantum
Monte Carlo method for general non-local in space and time
interactions. We demonstrated that for a Hubbard-type models the
computational time for a single trial step scales similarly to
that for the schemes based on a Stratonovich transformation. An
important difference occurs however for the non-local
interactions. Consider, for example, a system with a large Hubbard
$U$ and much smaller but still important Coulomb interatomic
interaction. One needs to introduce $N^2$ auxiliary fields per
time slice instead of $N$ to take the long-range forces into
account. On the other hand, the complexity of the present
algorithm should remain almost the same as for the local
interactions, because $\overline{|W|}$ does not change much. This
should be useful for the realistic cluster DMFT calculations and
for the applications to quantum chemistry \cite{qchem}. It is also
possible to study the interactions retarded in time, particularly
the super-exchange and the effects related to dissipation. This
was demonstrated for an important case of the fully rotationally
invariant two band model and its extension with non-local in time
spin-flip terms.

We are grateful to A. Georges, M. Katsnelson and F. Assaad for
their very valuable comments. This research was supported in part
by the National Science Foundation under Grant No. PHY99-07949,
"Russian Scientific Schools" Grant 96-1596476. Authors would like
to acknowledge a hospitality of KITP at Santa Barbara University
and (AR) University of Nijmegen.
%

%
%
%

\end{document}